\documentclass[useAMS,usenatbib]{mn2e}
\usepackage{txfonts}
\usepackage{multirow}
\usepackage{graphicx}

\usepackage{xcolor}
\usepackage{epsfig,graphics}
\usepackage{rotating} 
\usepackage{savefnmark}
\newcommand {\be}{\begin{equation}}
\newcommand {\ee}{\end{equation}}  
\renewcommand{\theenumi}{\arabic{enumi}.}

\title[Tidal disruption events in RASS-BSC]{Stellar tidal disruption
  candidates found by cross-correlating the \textit{ROSAT} Bright
  Source Catalogue and \textit{XMM-Newton} observations}   
\author[Khabibullin \& Sazonov]{I. Khabibullin$^{1,2}$\thanks{E-mail:
khabibullin@iki.rssi.ru}, S. Sazonov$^{1,3}$ \\
$^{1}$Space Research Institute, Russian Academy of Sciences,
Profsoyuznaya 84/32, 117997 Moscow, Russia\\
$^{2}$Max-Planck-Institut f\"ur Astrophysik,
Karl-Schwarzschild-Str. 1, 85740 Garching bei M\"unchen, Germany\\ 
$^{3}$Moscow Institute of Physics and Technology, Institutsky
per. 9, 141700 Dolgoprudny, Russia
}
\date{Received \today}

\begin{document}

\maketitle

\begin{abstract}

We performed a systematic search for stellar tidal disruption events
(TDE) by looking for X-ray sources that were detected during the
\textit{ROSAT} All Sky Survey and faded by more than an order of
magnitude over the next two decades according to \textit{XMM-Newton}
serendipitous observations. Besides a number of highly variable
persistent X-ray sources (like active galactic nuclei and cataclysmic
variables), we found \textit{three} sources that are broadly
consistent with the TDE scenario:  1RXS J114727.1+494302, 1RXS
J130547.2+641252, and 1RXS J235424.5-102053. A TDE association is also
acceptable for the fourth source, 1RXS J112312.7+012858, but an AGN
origin cannot be ruled out either. This statistics implies a TDE rate
of $ \sim 3\times 10^{-5} $ yr$ ^{-1} $ per galaxy in the Universe
within $z\sim 0.18$, which is broadly consistent with the estimates of
the TDE rate in the more local Universe obtained previously.
\end{abstract}


%%%%%%%%%%%%%%%%%%%%%%%%%%%%%%%%%%%%%%%%
\section{Introduction}
\label{s:intro}
%%%%%%%%%%%%%%%%%%%%%%%%%%%%%%%%%%%%%%%%

Tidal disruption of a star by a supermassive black hole (SMBH) is
expected to be a rather rare event, taking place every $ 10^3 - 10^6 $
years in a typical galaxy in the local Universe
\citep{Wang2004}. According to the canonical picture, some fraction ($
\sim 10\% $, \citealt{Ayal2000}) of the disrupted star's material
should be captured and then accreted by the SMBH on a time-scale of
about a year
\citep{Gurzadian1981,Rees1988,Evans1989,Phinney1989,Ulmer1999}. This 
results in the formation of an accretion disc emitting quasi-thermally
with the peak luminosity of $10^{43}-10^{45}$ erg s$^{-1} $, mainly at
extreme ultraviolet (EUV)/soft X-ray wavelengths
\citep{Strubbe2009}. Thus, a previously EUV/X-ray faint galaxy can 
become active for some period of time, whilst the stellar debris are
being accreted onto the SMBH, and then return to its initial quiescent 
state. Such tidal disruption events (TDE) can be searched for using
repeated wide-area surveys spaced in time by at least a year
\citep{Sembay1993,Komossa2002,Gezari2009,
  vanVelzen2011,Stone2013,Khabibullin2014}. 
 
Only about a dozen TDE candidates have been identified so far (see
\citealt{Komossa2012,Gezari2012} for recent reviews), with the major
contribution coming from X-ray observations
\citep{Komossa2002,Donley2002,Esquej2008,Cappelluti2009,
  Maksym2010,Lin2011,Saxton2012,Maksym2013}. The \textit{ROSAT}
All-Sky Survey (RASS, \citealt{Voges1999}) played the key role in the
first X-ray identifications and provided a reference point for
comparison with consequent \textit{ROSAT} pointed observations
\citep{Komossa2002,Donley2002} and later observations by
\textit{Chandra} and \textit{XMM-Newton}
(e.g. \citealt{Esquej2008}). Although up to several thousand TDEs had
been expected to be detected during the RASS \citep{Sembay1993}, only
\textit{five} were actually identified due to the lack of deep
archival and timely follow-up X-ray observations needed for probing
the pre- and/or post-outburst state
\citep{Komossa2002,Donley2002}. Therefore, plenty of TDEs will likely
remain hidden in the RASS data at least until the forthcoming X-ray
all-sky survey by the eROSITA telescope of the \textit{SRG} 
observatory (eRASS, \citealt{Khabibullin2014}).

Nevertheless, currently operating X-ray observatories have already
covered a significant fraction of the sky with sensitivity at least
ten times better than that of the RASS ($\sim 3\times
10^{-13}$~erg~s$^{-1}$~cm$^{-2}$ in the 0.5--2 keV energy band,
\citealt{Brandt2005}), thus providing an opportunity to search for
TDEs that occurred during the RASS epoch (in 1990--1991) and
have decayed since then. In the present study, we check all sources
from the RASS Bright Source Catalogue (RASS-BSC, \citealt{Voges1999})
for a large (more than a factor of ten) flux decrease in
  \textit{XMM-Newton} observations carried out between February 2000 and
  December 2012, which have covered $\sim 2$\% of the sky with a
  characteristic detection limit of
  $\sim10^{-14}$~erg~s$^{-1}$~cm$^{-2}$ in the 0.5--2 keV energy band.

Similarly to the present work, \cite{Donley2002} looked for X-ray
sources that occurred during the RASS. The difference is that we use
the much deeper \textit{XMM-Newton} serendipitous survey instead of
\textit{ROSAT} pointed observations for comparison with the RASS
data. This allows us to search for fainter TDEs, down to the flux
limit of RASS-BSC, in the fields observed by \textit{XMM-Newton}. As a
result, although the total area covered by \textit{XMM-Newton}
observations is much smaller than that covered by \textit{ROSAT}
pointed observations, we can probe the TDE rate out to a larger
distance ($z\lesssim 0.2$) compared to \cite{Donley2002} while
covering a similar volume of the Universe. Comparison of RASS and
\textit{XMM-Newton} data has previously been done by
\cite{Esquej2008}. However, their study was based on the
\textit{XMM-Newton} Slew Survey, whose sensitivity and sky coverage
are similar to the survey constructed from \textit{ROSAT} pointed
observations, and it was aimed at flares that occurred during the
\textit{XMM-Newton} epoch and were undetected by the RASS. Due to the
relatively small volume of the Universe probed by that study, it
discovered only two TDE candidates, which nonetheless proved to be of
high value thanks to their timely follow-up observations enabled by the
accurate \textit{XMM-Newton} localisations
\citep{Esquej2008}. 

In summary, with this study we aim to significantly increase the
available sample of TDE candidates and to begin to find TDEs outside
the local ($z<0.1$) Universe. In the latter regard, the present study 
is similar to the much larger forthcoming \textit{eROSITA} survey
\citep{Khabibullin2014}. An obvious drawback of the present study
is the impossibility of follow-up observations of TDEs detected during
the RASS due to their poor localisation and disappearance from the sky
by the present day.

The outline of the paper is as follows. In Section 2 we review the
databases and methods used in our analysis. A summary of the
identified TDE candidates is presented in Section 3. In Section 4 we
discuss the implications of this study for the TDE rate in
the Universe. Some conclusions are provided in Section 5.
  
%%%%%%%%%%%%%%%%%%%%%%%%%%%%%%%%%%%%%%%%
\section{Method}
\label{s:method}
%%%%%%%%%%%%%%%%%%%%%%%%%%%%%%%%%%%%%%%%

Our study follows a number of previous searches for X-ray sources
demonstrating a large amplitude flux drop ($>10$) and a 
soft, approximately $\sim 0.05$ keV blackbody, spectrum expected of TDEs
\citep{Komossa2002,Donley2002,Esquej2008,Cappelluti2009,Maksym2010,Lin2011,Saxton2012,Maksym2013}. Namely,
we looked for sources that were bright in the RASS epoch and faded
away over the next $\sim $10--20 years, as revealed by 
serendipitous observations with \textit{XMM-Newton}, since we do not 
expect any significant TDE-related emission at this very late phase.  

%------------------------------------------------------------------------------------------------- 

\subsection{ROSAT Bright Source Catalogue}
\label{ss:RBSC}

The \textit{ROSAT} All-Sky Survey Bright Source Catalogue (RASS-BSC)
was constructed from the all-sky survey performed during the first six
months of the \textit{ROSAT} mission in 1990/91 \citep{Voges1999}. It
contains 18,811 sources with a limiting count rate of 0.05 cts/s in
the 0.1--2.4 keV energy band, which corresponds to at least 15 photons
from a source. The typical positional accuracy is 30 arcsec
and at a brightness limit of 0.1 cts/s (8,547 sources), the catalogue
provides a sky coverage of 92\%. The catalogue also provides the source
extent, i.e. the amount by which the image of a source exceeds the
point spread function, which could be used to distinguish extended and
point sources.

The total energy band is divided in four channels: A($ \sim $0.1--0.4
keV), B($ \sim $0.5--2.0 keV), C($ \sim $0.5--0.9 keV) and D($ \sim
$0.9--2.0 keV), and spectral characteristics are presented in the form 
of hardness ratios HR1=(CR(B)-CR(A))/(CR(B)+CR(A)) and  
HR2=(CR(D)-CR(C))/(CR(D)+CR(C)), where CR(X) is the count rate in a
given channel X. 

There are two standard methods for RASS-BSC \citep{Voges1999} to
convert measured count rates to the unabsorbed 0.1--2.4 keV flux,
which differ in the assumptions about the source spectral shape. One
assumes an absorbed power law with a  
fixed photon index $\Gamma=2.3$, which is the typical value derived
from \textit{ROSAT} observations of extragalactic sources, and the 
absorbing column density fixed at the Galactic value along the given
line of sight (Flux 1). The other approach (Flux 2) is based on the
empirical relation for count rates and fluxes originally suited for
stars \citep{Voges1999}. For these methods, the accepted detection
limit of 0.05 cts/s translates to the following 0.1--2.4 keV flux
limits for point sources: $Flux1\sim 7\times
10^{-13}$~erg~s$^{-1}$~cm$^{-2}$ (assuming $ N_H=10^{20}$~cm$^{-2}$) and
$Flux2\sim 2\times 10^{-13}$~erg~s$^{-1}$~cm$^{-2}$ (for HR1=-1, see
the conversion formula in Section 3.3.5 of \citealt{Voges1999}). 

We make use of the FITS file with the catalogue available at the
anonymous ftp-server ftp.xray.mpe.mpg.de\footnote{Or directly at
  ftp://ftp.xray.mpe.mpg.de/rosat/catalogues/rass-bsc/}. The analysis
was performed using standard general purpose tasks for the
manipulation of FITS data contained in the the sub-package FUTILS of
the HEASARC's FTOOLS
package\footnote{http://heasarc.gsfc.nasa.gov/ftools/}
\citep{Blackburn1995}. 
%-------------------------------------------------------------------------------------------------
%-------------------------------------------------------------------------------------------------
\subsection{XMM-Newton Serendipitous Source Catalogue}
\label{ss:XMM}

3XMM-DR4\footnote{http://xmmssc-www.star.le.ac.uk/Catalogue/xcat\_public\_3XMM-DR4.html}  
is the third generation catalogue of X-ray sources serendipitously
detected by the European Space Agency's \textit{XMM-Newton}
observatory. It follows the previous 2XMM catalogue
\citep{Watson2009,Watson2013}.

The current version contains source detections drawn from 7427
\textit{XMM-Newton} EPIC observations made between 2000 February 3 and
2012 December 8, covering a total of 794 deg$^2$ of the sky,
accounting for overlaps. The position for a typical source can be
determined with $ \sim 2 $ arcsec accuracy \citep{Watson2009}. The
median flux in the soft (0.2--2 keV) energy band is $ \sim 6 \times
10^{-15}$~erg~s$^{-1}$~cm$^{-2}$, which is $ \sim 100 $ times lower
than the detection threshold accepted for RASS-BSC (see above). 

In this study we are interested not only in X-ray sources
  detected by \textit{XMM-Newton} but also in X-ray flux limits
  provided by \textit{XMM-Newton} observations. Hence, we use the XMM flux
  upper limit server
  (FLIX\footnote{http://www.ledas.ac.uk/flix/flix3}), which makes it 
  possible, by scanning the \textit{XMM-Newton} images corresponding 
  to the 3XMM-DR4 catalogue, to estimate an upper limit on the X-ray
  flux from a given position as well as to measure the flux within a
  circle centred on that position.

%-------------------------------------------------------------------------------------------------
%-------------------------------------------------------------------------------------------------
\subsection{Revealing decay of RASS-BSC sources with XMM-Newton
  observations} 
\label{ss:RXcross}
%-------------------------------------------------------------------------------------------------

Our goal is to obtain a sample of TDE candidates by cross-correlating
the RASS-BSC source catalogue with the \textit{XMM-Newton} data. As
usual in such studies, we should take both reliability and completeness
of the resulting sample into account. We prefer to use a more reliable
selection algorithm, since it is virtually impossible to perform
follow-up observations to confirm or reject a TDE association for RASS 
sources.

As a first step, we leave only sources outside the Galactic plane, by
requiring $ |b|>30^{\circ} $. There are two reasons for applying this
constraint. First, the Galactic extinction at $|b|<30^{\circ}$ is
expected to significantly diminish the fluxes of TDE flares due to
their soft X-ray spectra. Second, the increasing density of foreground
Galactic sources (primarily cataclysmic variables) can cause
significant contamination of the sample. However, the contribution of
Galactic contaminant sources remains noticeable ($\sim $50\%
 {of our raw candidate sample}, see below) even for $
|b|>30^{\circ} $. In addition, we exclude sources with the angular
extent larger than 30~arcsec, which allows us to avoid dealing with
extended sources like galaxy clusters. Finally, we produce a list of
positions that can be used as an input for the XMM flux upper limit
server. 

At the second stage, the XMM flux upper limit server reads the
file produced previously and returns a FITS-file with fluxes in
various energy bands measured within a circle centred on the RASS
position of a given source, provided there are \textit{XMM-Newton} 
observations of the corresponding field. We use the 0.2--2 keV energy
band and an extraction radius of 60 arcsec, which exceeds the RASS
localisation uncertainty, to obtain a conservative estimate. It is
important to note that such a flux estimate will obviously include
contributions from the background emission and from any sources
falling into the extraction region. At energies above 1 keV, 
it is dominated by the nearly constant contribution of extragalactic
sources with the total surface brightness of 4$ \times 10^{-15}  $ erg 
s$ ^{-1} $ cm$  ^{-2} $ arcmin $ ^{-2} $, while below 1 keV it is
dominated by diffuse Galactic and local thermal-like emission, 
which varies from position to position and has the net surface
brightness of the same order \citep{Hickox2006}. Hence, we can miss
potential TDE candidates located in highly crowded regions or regions
polluted by extended sources. In addition, we remove
\textit{XMM-Newton} observations with very small exposure time and
hence with a high uncertainty of the flux estimate. 

Finally, we check whether the flux derived from the
\textit{XMM-Newton} observations is at least 10 times lower than the
lowest of flux estimates provided by RASS-BSC (i.e. Flux1 or Flux2).
This resulted in a preliminary sample of 24 TDE candidates,
selected based on X-ray variability properties only. Since there are
only $\sim 20$ photons detected by \textit{ROSAT} for each of these
objects, we do not impose any additional constraints on our sample
based on X-ray spectral properties of the sources (reflected in their
hardness ratios, see Section \ref{ss:RBSC}). 

 { We then additionally examined these candidate sources
  for any problems that might have significantly affected their
  positions and fluxes reported in RASS-BSC. To this end, we
  first checked for the presence of cautionary flags in RASS-BSC,
  which may reflect some difficulties met by the detection
  algorithm. In addition, we visually compared available RASS and
  \textit{XMM-Newton} X-ray images for our candidates.}

 {
 As a result, one candidate (1RXS J132654.5-271104), which is reported
 as displaying a complex diffuse emission pattern ('d'-flag) but
 nevertheless considered point-like in RASS-BSC, is likely associated
 with the Abell 1736 galaxy cluster. The flux reported
 in RASS-BSC is collected from a much larger region (roughly,
 from the extraction region with a radius of 5 arcmin) compared to
 the region (with a radius of 1 arcmin) we used to get an upper limit
 from \textit{XMM-Newton} and can be dominated by X-ray emission from
 the intracluster medium.}

 {
 A somewhat different situation takes place in two other cases, 1RXS
 J003406.7-020935 and 1RXS J012605.2-012151, where the regions of
 interest appear to be contaminated by both a relatively large number
 of point sources and relatively weak diffuse emission from RXC
 J0034.6-208 (a galaxy cluster candidate from
 \citealt{Boehringer2004}) and Abell 194, respectively, as is clearly
 seen on \textit{XMM-Newton} images (see also \citealt{Hudaverdi2006}
 reporting the results of \textit{XMM-Newton} data analysis for Abell
 194). In both cases, one of the brightest point sources in the field
 falls rather close to the centroid provided by RASS-BSC, albeit well
 outside the cited localisation uncertainty region. Namely, it is 3XMM
 J003410.7-021039 (IC 0029), with a 0.2--12 keV flux of 2.8$ \pm $0.5
 $ \times 10^{-13}$ erg/s/cm$ ^2$, 88 arcseconds away from the RASS-BSC
 centroid for 1RXS J003406.7-020935 (having 1$ \sigma$ localisation
 error of 15 arcseconds), and 3XMM J012600.6-012041 (the NGC547/NGC545
 pair), with a 0.2--12 keV flux of 6.0$ \pm $0.2 $ \times 10^{-13}$
 erg/s/cm$ ^2$, 98 arcseconds apart from the RASS-BSC centroid for
 1RXS J012605.2-012151 (having 1$ \sigma$ localisation error of 25
 arcseconds). Assuming that it is these sources which are in fact
 responsible for the fluxes reported in RASS-BSC, these candidates
 then fail to satisfy our TDE selection criterion based on high-amplitude
 variability. } 

 { 
Thus, there remain 21 candidates in our sample. However, some of them
still have cautionary RASS-BSC flags due to the presence of a nearby
bright source, and we will discuss this specifically in every
interesting case below.}  
 
%-------------------------------------------------------------------------------------------------
\subsection{Cross-correlation with other surveys}
\label{ss:cross}
%-------------------------------------------------------------------------------------------------

Since many of the RASS-BSC sources have been identified and classified
since they were discovered, 
%\citep{Voges1996,Zickgraf1997,Appenzeller1998,Bade1998,Thomas1998,
%Beuermann1999,Bauer2000,Rutledge2000,Zimmermann2001,Anderson2003,
%McGlynn2004,Anderson2007,Agueros2009,Haakonsen2009,Mahony2010} 
we searched NASA/IPAC Extragalactic
Database\footnote{http://ned.ipac.caltech.edu/} (NED) and the
SIMBAD Astronomical
  Database\footnote{http://simbad.u-strasbg.fr/simbad/} for possible
identifications of our TDE candidates. \textit{Nine} of the candidates
turned out to be known stellar-type objects, including 7 systems with
accreting white dwarfs (RX J0132.7-6554, \citealt{Burwitz1997}, EF
Eri, \citealt{Verbunt1997}, RX J0527.8-6954,
\citealt{Truemper1991,Greiner1996}, RX J1039.7-0507,
\citealt{Appenzeller1998}, WGA J1047.1+6335, \citealt{Singh1995}, RX
J1957.1-5738, \citealt{Thomas1996}, RX J2022.6-3954,
\citealt{Burwitz1997}), a symbiotic star, CD--43$^\circ$14304
\citep{Muerset1997}, and a hot white dwarf, WD~J2324-546
\citep{Pounds1993}. 
 
 \textit{Four} other candidates have been reported to demonstrate
 signatures of the active galactic nucleus (AGN) activity: KUG
 1624+351 -- a Seyfert 1.5 (Sy1.5) galaxy at $z= 0.0342$
 \citep{Bade1998,Veron2006}, RX J1225.7+2055 -- a narrow line Seyfert~1
 (NLSy1) galaxy at $z= 0.335$ \citep{Greiner1996b}, MCG -01-05-031 --
 a Seyfert~2 (Sy2) galaxy at $z=0.0182$ \citep{Panessa2002} and WPVS
 007 -- a NLSy1 galaxy at $z=0.028$. The last object is famous for
 very unusual X-ray properties, primarily soft X-ray variability by
 two orders of magnitude over three years \citep{Grupe1995} and has
 already been mentioned in the context of TDE searches
 \citep{Donley2002}. However, the very recent extensive monitoring of
 WPVS 007 by the \textit{Swift} observatory clearly indicates a 
 high level of short-term variability of the source consistent with a
 partial covering absorber model \citep{Grupe2013}. Having been
 observed by \textit{XMM-Newton} for $\sim 100$ kiloseconds (2010 June
 11), the source was identified as 3XMM J003915.8-511701 with the mean
 0.2--12 keV flux of $(1.12 \pm 0.16) \times 10^{-14}
 $~erg~s$^{-1}$~cm$^{-2}$, i.e. $\sim 500$ times fainter compared with
 the detection during the RASS. The other three sources have also been
 marginally detected by \textit{XMM-Newton} and Table \ref{t:agn}
 provides brief information about these detections.
 
  \begin{table*}
\caption{Strongly variable X-ray sources with AGN signatures}
\begin{tabular}{cccccccccc}
\hline\hline

\multirow{2}*{Name} & \multirow{2}*{Type$^a$}&\multirow{2}*{z$^a$}&
\multicolumn{3}{c}{3XMM-DR4}&\multicolumn{3}{c}{RASS-BSC}
&\multirow{2}*{Amplitude$^{d}$}\\  

& & & 3XMM ID & Date (UTC) & Flux$^{b}$& Count
rate$^{c}$ & Flux1$^{c}$ &Flux2$^{c}$&\\ 
\hline
WPVS 007 & NLSy1 &0.028  & J003915.8-511701  & 2010-06-11.9	&1.12
$ \pm $ 0.16 &0.96$ \pm $0.068 &19.4&3.04&$ > 270 $	\\ 
MCG -01-05-031 & Sy2 &0.0182  & J014525.4-034938  & 2008-08-03.2
&5.97$ \pm $0.7 &0.17$ \pm $0.03 &3.5&0.92	&$ > 15$	\\ 
RX J1225.7+2055 & NLSy1 &0.335  &J122541.9+205503 &2003-06-12.7	&9.95$
\pm $ 1.19&0.33$ \pm $0.03 &6.7&1.63&$ > 16$	\\ 
\multirow{2}*{KUG 1624+351} & \multirow{2}*{Sy1.5} &\multirow{2}*{
  0.0342}  &\multirow{2}*{J162636.5+350241}	& 2007-08-17.1& 3.685$
\pm $ 0.708&\multirow{2}*{ 0.081$ \pm $0.013}
&\multirow{2}*{0.971}&\multirow{2}*{0.672}	&$ > 18 $	\\ 
& & & &2007-08-19.2 & 3.61$ \pm $ 0.59 & &&	 &$ > 18 $ \\
\hline
\end{tabular}
\\
\begin{flushleft}
\footnotesize{$ ^a $ See references in the text.\\
$ ^{b} $ The 0.2--12 keV flux as measured by
  \textit{XMM-Newton}, 
  $10^{-14}$~erg~s$^{-1}$~cm$^{-2}$. \\    
$ ^{c} $  \textit{ROSAT} count rate with a 1$\sigma$ uncertainty
  (cts/s) and corresponding estimates for the 0.1--2.4 keV flux (Flux1
  and Flux2, see text), $ 10^{-12} $~erg~s$^{-1}$~cm$^{-2}$.\\  
$ ^{d} $ Estimated amplitude of the flux drop between the RASS-BSC and
3XMM-DR4 epochs. It is calculated relative to the lowest of Flux1 and
Flux2, so the actual amplitude should be somewhat higher.} 
\end{flushleft}
\label{t:agn}
\end{table*}
 
The remaining \textit{eight} candidates (see Table~\ref{t:clean} and
Table~\ref{t:cleanxmm}) either do not have firm identification or are
associated with galaxies with no obvious AGN signatures. We have
looked for potential counterparts  of these objects inside their
\textit{ROSAT} localisation regions using the Guide Star Catalog II
(GSC-II)\footnote{http://archive.eso.org/gsc/gsc} based on the
  Digitized Sky Survey (DSS) images \citep{Lasker2008}, the Two Micron
  All Sky Survey (2MASS,
  \citealt{Skrutskie2006})\footnote{http://www.ipac.caltech.edu/2mass},
  the \textit{Wide-field Infrared Survey Explorer} (\textit{WISE})
  all-sky
  survey\footnote{http://irsa.ipac.caltech.edu/Missions/wise.html} 
\citep{Wright2010} and the Sloan Digital Sky Survey
(SDSS)\footnote{http://www.sdss3.org/}. The next section presents the
results of this search\footnote{The cross-correlation analysis was
  conducted mainly using the TOPCAT software
  \citep{Taylor2005}.}.
 
 \begin{table*}
 \caption{RASS-BSC data for potential TDE candidates}
 \begin{tabular}{cccccccccccccc}\hline\hline
1RXS ID & RA & Dec & PosErr$ ^a $ & CR$ ^{b} $ & eCR$ ^{b} $ & Exp$ ^{b} $ & HR1$ ^{c} $ & eHR1$ ^{c} $ & HR2$ ^{c} $ & eHR2$ ^{c} $ & Flux1$ ^{d} $ & Flux2$ ^{d} $ &N$_H^{e} $\\\hline
J002048.5-253823 & 5.20208 & -25.63986 & 16(26) & 0.0563 & 0.0167  & 280 & -0.32 & 0.27 & -0.82 & 0.87 & 9.78 & 3.72 & 2.3\\
J005626.3-010615 & 14.10958 & -1.10431 & 15(24) & 0.0999 & 0.0202 &  312 & 0.65 & 0.18 & 0.61 & 0.19 & 21.5 & 11.7& 3.4 \\
J101326.2+061202 & 153.35918 & 6.20069 & 15(25) & 0.0618 & 0.0154 &  434 & -1 & 0.13 & 0 & 0 & 11.6 & 3.7& 2.0\\
J112312.7+012858 & 170.80292 & 1.48292 & 17(28) & 0.052 & 0.0143 &  375 & 0.08 & 0.27 & 0.15 & 0.35 & 13.7 & 4.53&3.5\\
J114727.1+494302 & 176.86292 & 49.71722 & 9(15) & 0.414 & 0.0402 &  299 & -0.97 & 0.02 & 1 & 1.3 & 63.3 & 13.1&1.7\\
J130547.2+641252 & 196.44666 & 64.21445 & 9(15) & 0.167 & 0.0207 &  544 & -0.82 & 0.06 & -0.62 & 0.28 & 23.2 & 6.61&1.5\\
J215101.5-302852 & 327.75626 & -30.48111 & 14(23) & 0.0534 & 0.0159 &  333 & -0.34 & 0.24 & 0.42 & 0.47 & 8.29 & 3.47&1.7\\
J235424.5-102053 & 358.60208 & -10.34806 & 24(39)  & 0.0523 & 0.0153 &  310 & 1 & 0.25 & -0.13 & 0.29 & 10.8 & 7.12&2.7\\
\hline
\label{t:clean}
\end{tabular}
\\
\begin{flushleft}
\footnotesize{$ ^a$ 1$ \sigma $ localisation uncertainty and
  corresponding radius of the 95\% confidence region, arcsec.\\ 
$ ^{b} $ The source mean count rate and corresponding 1$ \sigma $
  error (both in units of counts~s$^{-1}$) and exposure time (in sec).\\
$ ^{c} $ Hardness ratios (see text) with corresponding 1$ \sigma $
  errors.\\  
$ ^{d} $ 0.2--2.4 keV flux estimates (see text), 10$^{-13} $ erg
  s$^{-1} $ cm$ ^{-2} $. \\ 
$ ^{e} $ Total Galactic H I Column Density (divided by $10^{20} $ cm$
  ^{-2}  $) as determined with the $nH$ utility of the $ FTOOLS $
  package (based on the map of \citealt{Kalberla2005}).\\ 
}
\end{flushleft}
 \end{table*}

%-------------------------------------------------------------------------------------------------
%%%%%%%%%%%%%%%%%%%%%%%%%%%%%%%%%%%%%%%%
\section{Results}
\label{s:results}					
%%%%%%%%%%%%%%%%%%%%%%%%%%%%%%%%%%%%%%%%

Table~\ref{t:clean} summarizes the RASS-BSC data for the potential TDE
candidates. Table~\ref{t:cleanxmm} provides X-ray flux limits
  for these sources obtained more than 10 years later with
  \textit{XMM-Newton}. Since the typical RASS localisation
uncertainty is $r_{1\sigma} \sim 15$ arcsec, it is difficult to
uniquely associate these TDE candidates with sources from IR, optical
or high angular-resolution X-ray surveys, and we should regard all
sources (if any) falling into the RASS localisation region as possible
counterparts. We use the $ 2\sigma $ (i.e. $ \approx 95\% $)
confidence region (corresponding to a radius $r_{2\sigma} \approx 1.6
r_{1\sigma} $ for the two-dimensional Gaussian distribution) for such
analysis. Hence, only a few per cent of true counterparts can be missed.

Typically, there are several possible counterparts in the RASS localisation
region. For example, this is the case if SDSS photometric data are
available. The easiest possibility of proceeding with identification
in such cases is checking for sources with AGN or cataclysmic variable
(CV) signatures and
thus rejecting a TDE origin. This can be done in several ways. First,
a TDE flare taking place in the RASS epoch is expected to disappear by
the time of the \textit{XMM-Newton} observations, while the X-ray
emission from the TDE host (inactive) galaxy is very unlikely to 
exceed the \textit{XMM-Newton} detection limit. In contrast, there can
well be a significant X-ray detection in 3XMM-DR4 (of course, this
flux being at least 10 times lower than the flux measured by
\textit{ROSAT}) in the case of an AGN or CV. Second, any spectral
information could of course be helpful but, unfortunately, an optical
spectrum is available only for a small fraction of the possible
counterparts. Finally, WISE infrared (IR) colors can also  indicate an
AGN origin, and such information is indeed available for many of the
possible counterparts. 

In addition, we can put a simple, physically motivated constraint on
the expected optical brightness of a TDE host galaxy. Indeed, 
given the mass of the central SMBH, $M_{BH}$, the (near) peak soft
X-ray luminosity of a TDE is expected to be 
\begin{equation}
L_{X}=kL_{Edd}(M_{BH})=kL_{Edd,\odot}\frac{M_{BH}}{M_{\odot}},
\end{equation}
where $L_{Edd,\odot}\simeq 1.4\times 10^{38}$~erg/s is the Eddington
luminosity for one solar mass and the factor $ k \sim 0.1$ accounts
for the bolometric correction, possible geometrical dilution and the
fact that the X-ray flux detected by RASS may be measured not exactly
at the peak of the TDE (see e.g. \citealt{Khabibullin2014}). 
 
On the other hand, the optical luminosity of the TDE host galaxy could
be roughly expressed as
\begin{equation}
L_{V}\simeq \frac{L_{V,bulge}}{P_b}\simeq
\frac{200}{P_b}L_{V,\odot}\frac{M_{BH}}{M_\odot},
\label{eq:Lv}
\end{equation}
where $ L_{V,\odot}\simeq 4.6\times 10^{32}$~erg/s is the solar
V-band luminosity, $ P_b \sim 0.1 $ is the fraction of the bulge
in the total galaxy luminosity\footnote{This value for $ P_b $ is
  typical for disk galaxies but not for elliptical ones. However, for
  elliptical galaxies $ M_{BH} $ typically exceeds the tidal
  disruption threshold for solar-type stars ($ \sim 10^8 M_\odot $),
  so we do not expect significant contribution from such
  galaxies in our TDE sample.}, and we have converted $L_{V, bulge}$ to
$M_{BH}$ using the $M_{BH}-L$ relation from \cite{Gultekin2009} for $
M_{BH}\sim 10^7 M_{\odot} $  {(in fact, the $M_{BH}-L$
  relation is almost linear, so the numerical factor in
  Eq.~\ref{eq:Lv} depends on $ M_{BH}$ very weakly).}

As a result, we obtain the following relation between the TDE soft
X-ray ($\sim$0.2--2~keV) flux, $F_{X}$, and the optical flux of the
host galaxy, $F_V$:
\begin{equation}
F_{V}=F_{X}\frac{L_{V}}{L_{X}}\simeq \frac{F_{X}}{k~P_b}
\frac{200~L_{V,\odot}}{L_{Edd,\odot}}\simeq 0.064\frac{F_{X}}{k_1~P_{b,1}},
\end{equation}
where $ k_1=k/0.1 $ and $ P_{b,1}=P_b/0.1 $. 

For $F_{X}=3\times 10^{-13} $ erg/s/cm$ ^2 $, i.e. close to the
RASS-BSC detection limit, $F_V=1.9\times 10^{-14} $ erg/s/cm$^2$, 
given $ k=0.1 $ and $ P_{b}=0.1 $, which translates to
$m_{V}=2.5(-5.5-\log F_V)\simeq 20.55$. For
$F_{X}=1\times 10^{-12} $ erg/s/cm$ ^2 $, a typical flux for RASS-BSC
sources (and for our candidate sources, see Table~\ref{t:clean}), one
gets $ m_{V}\simeq 19.25$. 
 % AB magnitude for g-r SDSS filters. 
Thus, a TDE host galaxy is unlikely to be very faint, unless we deal
with an outlier from the $M_{BH}-L_{V,bulge}$ relation. We caution
though that the $M_{BH}-L_{V,bulge}$ correlation used above was
derived from a sample with somewhat heavier SMBHs compared to those
expected to produce most TDEs in the Universe.

Similar estimates can be done for the near-IR (e.g. K band) magnitude
of TDE host galaxies. According to the $M_{BH}-L_{K}$ relation for
the total (bulge+disk) K-band luminosity from
\cite{Lasker2014},
\begin{equation}
L_{K}\simeq {300} L_{K,\odot}\frac{M_{BH}}{M_\odot}
\label{eq:Lk}
\end{equation}
for $ M_{BH}\sim 10^{7} M_\odot $\footnote{ {In this
    case, the $M_{BH}-L_{K}$ relation is also almost linear, so the
    numerical factor in Eq.~\ref{eq:Lk} will be essentially the same for $
    M_{BH}\sim 10^6 M_\odot $ .}}, where $L_{K,\odot}\simeq 8\times 
10^{31}$~erg/s is the solar K-band luminosity
(e.g. \citealt{Binney1998}, which corresponds to a Vega magnitude
M$_K=3.28$). Therefore,   
\begin{equation}
F_{K}=\frac{F_{X}}{k}\frac{300~L_{K,\odot}}{L_{Edd,\odot}}\simeq
2\times10^{-3}\frac{F_{X}}{k_1}. 
\end{equation}
Hence, for $ F_{X}=1\times 10^{-12} $ erg/s/cm$ ^2 $, $F_{K}=2\times
10^{-15} $ erg/s/cm$ ^2 $, or $m_{K} \simeq 19.4 $\footnote{The
  flux-to-magnitude conversion was calculated as $m_K=2.5(-6.95-\log
  F_K)$, in accordance with the 2MASS photometric system
  \citep{Cohen2003}.}. 
 
For different classes of contaminating sources, yet simpler
calculations can be carried out. As known from X-ray surveys,
$0.1\lesssim F_X/F_V \lesssim 10$ for AGN and CVs in the soft X-ray 
band (e.g. \citealt{Maccacaro1988,Aird2010}), which translates to
$15.15 \lesssim m_V \lesssim 20.15$ given $ F_X=3\times10^{-13}
$~erg/s/cm$^2$. Somewhat higher $ F_X/F_V $ ratios could be found 
during flares of luminous blazars
\citep{Maccacaro1988,Cenko2012}. It should be
  mentioned however that in the case of AGN and CVs, optical
  counterpart emission can be variable and tightly
  related to the X-ray emission. So if we compare the (relatively
  high) X-ray flux measured during the RASS with non-contemporaneous
  optical observations (e.g. SDSS), the resulting X-ray/optical flux
  ratio may be different, likely higher, than estimated above, hence
  the optical counterpart may be somewhat fainter.

For flaring stars, $ F_X/F_V \sim 10^{-3}$--$10^{-2} $ is much more
typical (with rare exceptions of very intense flares in low-mass
stars, \citealt{Favata2003}\footnote{See also Section 6.5 in 
  \cite{Merloni2012} for a relevant discussion.}), so the optical
counterpart is expected to be brighter than $ m_V \simeq 13 $ in this
case.

In summary, the optical counterparts of our candidate sources are
likely to be present in the SDSS photometric data (complete to $r\sim
22.2$, \citealt{Abazajian2004}), if available, even in the case of
AGN/CV association, and they are probably sufficiently  
bright to be present in DSS images in the case of TDE association
(GSC-II is complete to $ R_F \simeq 20$, \citealt{Lasker2008}).

\begin{table*}
 \caption{XMM-Newton flux limits for potential TDE candidates}
\begin{tabular}{cccccccc}\hline\hline
1RXS & XMM ObsID & OffAxis$^a  $ & Date & Exposure, s & Flux limit$^{b}  $ &Ratio1$ ^{c} ~$&Ratio2$ ^{c} ~$\\\hline
\smallskip
{ J002048.5-253823} & 201900301 & 4.386714 & 2004-05-26 & 9160 & 2.3 $ \pm $ 0.3&43&	16\\
\multirow{3}*{J005626.3-010615} & 12440101 & 13.3243 & 2001-01-15 & 7711 & 3.1 $ \pm $ 0.8&70&	38\\
 & 402190501 & 8.833312 & 2006-06-16 & 6270 & 5.9 $ \pm $ 1.8&37&	20\\
\smallskip
 & 505211001 & 8.702417 & 2007-07-15 & 4905 & 3.6 $ \pm $ 1.0&60&	33\\
\smallskip
J101326.2+061202 & 600920301 & 4.105885 & 2009-05-26 & 4337 & 2.8 $ \pm $ 0.5&41&	13\\
\smallskip
J112312.7+012858 & 145750101 & 7.475069 & 2003-06-23 & 9966 & 5.4 $ \pm $ 0.4&25&	8.4\\
\smallskip
J114727.1+494302 & 604020101 & 1.638498 & 2009-11-21 & 1632 & 1.6 $ \pm $ 0.7&410&	84\\
\smallskip
J130547.2+641252 & 151790701 & 1.726202 & 2003-10-10 & 4456 & 0.4 $ \pm $ 0.9&560&	160\\
\smallskip
J215101.5-302852 & 103060401 & 10.47006 & 2001-05-01 & 7341 & 4.2 $ \pm $ 0.6&20&	8.2\\
J235424.5-102053 & 108460301 & 5.601944 & 2001-06-20 & 11009 & 1.1 $ \pm $ 0.5&96&	63\\\hline
\end{tabular}
\begin{flushleft}
\footnotesize{$^a$ Offset of the candidate RASS-BSC source (the first
  column) from the centre of the \textit{XMM-Newton} field of view
  during a given observation (specified in the second column),
  arcmin. \\ 
$^{b}$ Estimated 0.2--2 keV flux within the 60 arcsec-radius region
  around the RASS-BSC position, with a $1\sigma$ uncertainty,
  $10^{-14}$~erg/s/cm$^2$. Note that this flux includes the
    background contribution. \\
$ ^{c}$ \textit{ROSAT/XMM-Newton} flux ratio, found by comparison with
the Flux1 or Flux2 estimate for RASS-BSC (see Table~\ref{t:clean}).} 
\end{flushleft}
 \label{t:cleanxmm}
 \end{table*} 

%%%%%%%%%%%%%%%%%%%%%%%%%%%%%%%%%%%%%%%%
\subsection{1RXS J002048.5-253823}
%-------------------------------------------------------------------------------------------------

 {
1RXS J002048.5-253823 is located approximately 5 arcminutes from
the galaxy cluster Abell 022, which is present in RASS-BSC as an
independent X-ray source, 1RXS J002041.8-254307, and is $ \sim $ 6
times brighter than 1RXS J002048.5-253823. Therefore, the localisation
and the flux estimate could be somewhat biased for 1RXS
J002048.5-253823. However, examination of the RASS image shows that 
the two sources are well separated from each other. The available
crude spectral information (hardness ratios), HR1=$-0.32\pm 0.27 $ and
HR2=$-0.82\pm0.87 $ for 1RXS J002048.5-253823 against HR1=$ -0.31\pm
0.14 $ and HR2=$-0.22\pm 0.15 $ for 1RXS J002041.8-254307, is of
little value in this respect given the large uncertainties for the
fainter source. 
}
 
\begin{figure}
\centering
\caption{The Digitized Sky Survey (DSS) image (2'x2') of the region
  around 1RXS~J002048.5-253823. The dashed and solid blue circles
  confine the $1\sigma$ and $2\sigma$ regions around the centroid of
  the \textit{ROSAT} source,  respectively. Red squares mark 2MASS
  sources, green crosses mark 3XMM sources (if any). For those fields
  with SDSS data available (see the subsequent figures), black
  diamonds mark the positions of SDSS sources. The arrow points out
  the most probable counterpart.} 
\includegraphics[width=1.2\columnwidth]{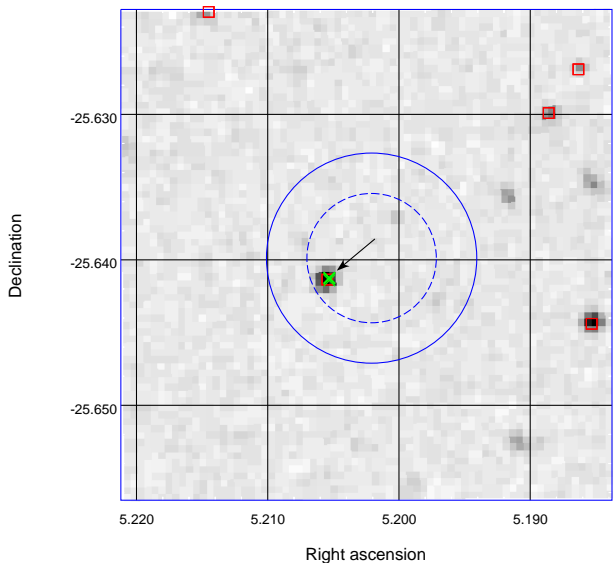}
\label{f:s1}
\end{figure}

Moreover, there is a faint 3XMM-DR4 source (3XMM J002049.3-253828) with the
0.2--2 keV flux of (1.7$ \pm  $0.5)$ \times 10^{-14}$ erg/s/cm$^2$,
i.e. close to the detection limit of the 3XMM-DR4 catalogue, inside
the $ 1\sigma $-region (12 arcsec away from the reported RASS-BSC
centroid). This source has an optical counterpart in the DSS image
(and the corresponding entry in GSC-II with $R_F=17.1$), a
near-infrared counterpart in the 2MASS catalogue and a mid-infrared
counterpart in the \textit{WISE} catalogue (see Fig.~\ref{f:s1} and
Table \ref{t:cross}). The \textit{WISE} colors (namely, $W2-W1=1.24 >
$0.8) strongly indicate an AGN association for this source
\citep{Stern2012}. Since there are no other objects inside the RASS
localisation region, we suggest that 1RXS~J002048.5-253823 is a
high-amplitude ($\gtrsim 50$ if Flux1 is used as a flux estimate for
the RASS-epoch) AGN flare. 

We conclude that 1RXS J002048.5-253823 is probably not a TDE. 

%%%%%%%%%%%%%%%%%%%%%%%%%%%%%%%%%%%%%%%%
\subsection{1RXS J005626.3-010615}

 {
This candidate is also located nearby a galaxy cluster, Abell 119,
though approximately 10 arcminutes from its centre. Abell 199
enters RASS-BSC as 1RXS J005617.5-011501, being $ \sim $ 5 times
brighter than 1RXS J005626.3-010615 with the extension of $ \gtrsim $
2 arcmin. So, the sources seem to be well separated for
the extraction region of the fainter one (10 arcmin in diameter)
to include almost no extended emission from the brighter one. Both
hardness ratios are also formally different for them: HR1=$0.65\pm
0.18  $, HR2=$0.61\pm 0.19 $ for 1RXS J005626.3-010615 against HR1=$
0.88\pm 0.05 $, HR2=$0.29\pm 0.08  $ for 1RXS J005617.5-011501. }   

\begin{figure}
\caption{DSS image (2'x2') of the region around 1RXS J005626.3-010615.
The symbols are the same as for Fig.~\ref{f:s1}.}
\includegraphics[width=1.2\columnwidth]{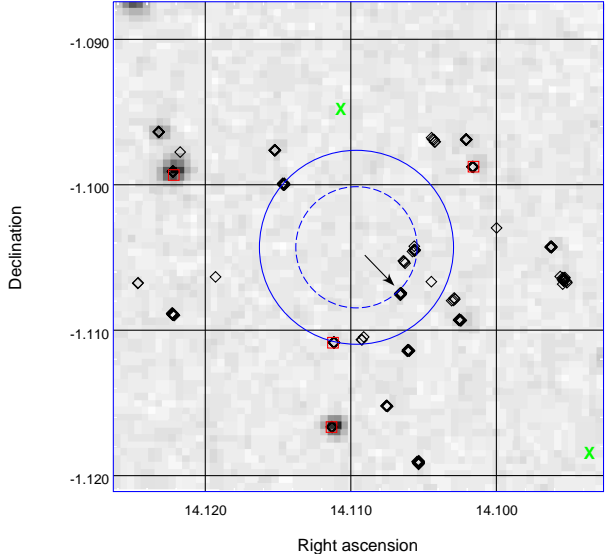}
\label{f:s3}
\end{figure}

There are seven SDSS sources inside the localisation region (see
Fig.~\ref{f:s3}). All of them are faint, $ r \gtrsim 20 $, and thus
invisible in the DSS image. An optical spectrum is available for only
one source, which can be unambiguously classified as an M-star. This
source (24~arcsec from the RASS centroid) is also the only one with a
2MASS counterpart. Although such a star could potentially produce an
intense X-ray flare, it should be much brighter since the X-ray flux
in the \textit{ROSAT} energy range exceeded $ 10^{-12} $ erg/s/cm$^2 $
and one should then expect $ m_{V}\lesssim 16.5 $ assuming $
F_X/F_V\lesssim 1 $ (such ratios correspond to the most extreme stellar
flares).

Three other sources are identified as extended ones and another three
are consistent with being point-like. The data on these sources are
given in Table~\ref{t:cross}. An extended source 16~arcsec away from
the RASS-BSC centroid is the only one detected by \textit{WISE} and
its colors favour an  AGN association. It is also the optically
brightest one with $ r\approx21.0 $. This object can thus be
considered the most probable counterpart. As to its nature, it could
be a highly variable AGN (perhaps a blazar) with a very high  $
F_X/F_V $ ratio.

We thus conclude that 1RXS J005626.3-010615 is unlikely to be a TDE. 

%%%%%%%%%%%%%%%%%%%%%%%%%%%%%%%%%%%%%%%%
\subsection{1RXS J101326.2+061202 }

\begin{figure}
\caption{DSS image (2'x2') of the region around 1RXS J101326.2+061202.
The symbols are the same as for Fig.~\ref{f:s1}.}
\includegraphics[width=1.2\columnwidth]{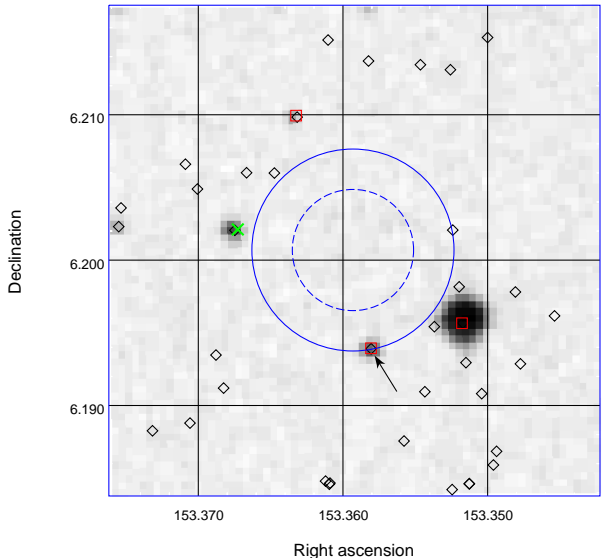}
\label{f:s5}
\end{figure}

There are two SDSS sources at the boundary of the localisation region,
25~arcsec away from the RASS-BSC centroid (see Fig.~\ref{f:s5}). One
is classified as a K star according to the SDSS 
spectrum, so it potentially could be responsible for the highly
variable X-ray emission. Indeed, the optical magnitudes of the source
satisfy the $ F_X/F_V \lesssim 1 $ condition discussed above. This
source is also the only one inside the uncertainty region with 2MASS
(with $ JHK $ magnitudes of 15.56, 14.86 and 14.82) and WISE
(with $W1=14.69$ and $W2=14.79$) counterparts. The other
source is too faint to even be considered a possible counterpart in
the case of a TDE association (see Table \ref{t:cross}). 
Thus, we suggest that the most probable counterpart is the K star,
which experienced a strong flare during the RASS. 

We conclude that 1RXS J101326.2+061202 is unlikely to be a TDE.

%%%%%%%%%%%%%%%%%%%%%%%%%%%%%%%%%%%%%%%%
\subsection{1RXS J112312.7+012858}

There is a 3XMM-DR4 source (3XMM J112313.1+012924), with a flux of $
(6.0\pm0.8)\times10^{-14}$ erg/s/cm$^2$ in the 0.2--2 keV energy band,
at the boundary of the 
localisation region (28 arcsec from the RASS-BSC centroid). This
source has a faint, non-extended SDSS counterpart with (22.2, 21.3,
21.0, 21.4, 20.8)  $ ugriz $ magnitudes, but there are no 2MASS or
\textit{WISE} counterparts. Taking these facts together, we consider
association of the 3XMM-DR4 source with 1RXS J112312.7+012858 unlikely. 

In total, there are eleven SDSS sources inside the localisation
region. One of these has been confidently identified as an A0 star by
means of optical spectroscopy. The data on all other sources are
summarized in Table \ref{t:cross}. The brightest SDSS object ($
r=19.9 $), 21 arcsec away from the RASS-BSC centroid, is consistent
with an extended source and is also the only one with a $ WISE $
counterpart. The first \textit{WISE} color
($W1-W2=15.114-14.831=0.283$) makes a luminous AGN association
unlikely for this source. However, this object can be the host galaxy  
of a TDE. 

We conclude that 1RXS J112312.7+012858 is possibly a TDE.

\begin{figure}
\caption{DSS image (2'x2') of the region around 1RXS J112312.7+012858} 
\includegraphics[width=1.2\columnwidth]{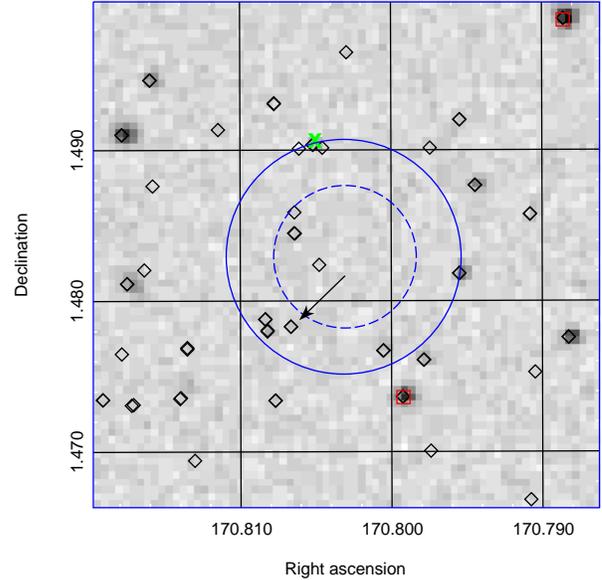}
\label{f:s6}
\end{figure}

%%%%%%%%%%%%%%%%%%%%%%%%%%%%%%%%%%%%%%%%
\subsection{1RXS J114727.1+494302}

This source, also known as RBS 1032, has been intensively investigated 
by \cite{Ghosh2006}. They reported an ultrasoft X-ray spectrum, consistent
with a blackbody with $ \kappa T_{bb}\simeq 70 $ eV, as well as a $ \sim $
3-fold flux drop from the original RASS detection
(November 05, 1990) to the first pointed \textit{ROSAT} observation
carried out $ \sim  $ 2.1 years later (December 07, 1992) and a $ \sim
$ 6-fold drop to the second pointed observation $ \sim $ 3.6 years
later (June 05, 1994) \citep{Ghosh2006}. The authors identified
RBS~1032 with a dwarf spheroidal galaxy (SDSS~J114726.69+494257.8) at
$z=0.026$ with no optical signatures of nuclear activity. The
\textit{WISE} colours of this source ($W1-W2=0.115$) also rule out a
luminous AGN. \textit{XMM-Newton} detected only a marginal signal from
this position in late 2009 with the net 0.2--12 keV flux of (1.1 $ \pm
$ 0.6) $ \times $10$ ^{-14} $ erg s$ ^{-1} $ cm$ ^{-2} $.

\begin{figure}
\caption{DSS image (2'x2') of the region around 1RXS J114727.1+494302}
\includegraphics[width=1.2\columnwidth]{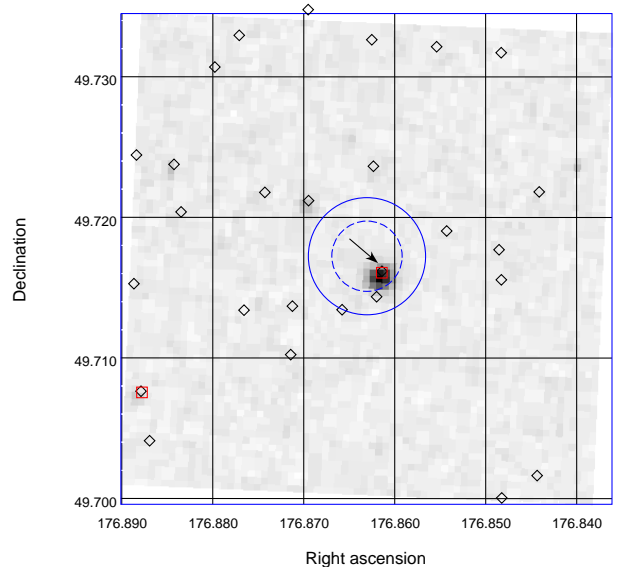}
\label{f:s7}
\end{figure}

There are two other SDSS extended sources inside the localisation
region, with neither 2MASS nor \textit{WISE} detection (see
Fig.~\ref{f:s7} and Table \ref{t:cross}). These sources are too faint
to be considered potential TDE hosts.  
 
\cite{Ghosh2006} argued that the variable X-ray emission may come from
an intermediate mass black hole accreting matter from a white
dwarf companion. The estimated soft X-ray luminosity at peak is $ \sim
10^{43} $ erg/s/cm$ ^2 $ and fitting the X-ray spectra with a
blackbody model yields a temperature of $\kappa T_{bb}\simeq 70$ eV
\citep{Ghosh2006}. Both of these characteristics are consistent 
with the TDE scenario \citep{Komossa2002,Esquej2008}. However, the
six-fold flux decrease in the 3.6 years spanned by the
\textit{ROSAT} observations may seem to be too small for the TDE scenario
assuming that the first detection occurred near the TDE peak. 

However, the estimated luminosity is not sufficiently high to exclude
the possibility that the first detection in fact occurred at a relatively
late phase of the event. Indeed, assuming the canonical shape of
the decay phase of the TDE light curve \citep{Phinney1989}, 
\begin{equation}
\frac{L(\tau +\delta t)}{L(\tau)}=\left(\frac{\tau +\delta
  t}{\tau}\right)^{-5/3}, 
\end{equation}
where $ \tau $ is the time of the RASS observation relative
to the disruption moment and $ \delta t =3.6 $ yrs is the time passed
between the RASS observation and the second pointed \textit{ROSAT}
observation. Given that the unabsorbed 0.1--2.4 keV flux decreased
from 6.0$ \times $10$ ^{-12} $ erg s$ ^{-1} $ cm$ ^{-2} $ to 1.1$
\times $10$ ^{-12} $ erg s$ ^{-1} $ cm$ ^{-2} $ during this period
\citep{Ghosh2006}, we find $L(\tau +\delta t)/L(\tau)=1.1/6.0 $, so $
\tau \simeq 2.0 $ yrs. This implies that the disruption took place
near the 5th of November, 1988. Since the duration of the early
(Eddington-limited) phase $ \tau_{Edd}\simeq 0.25~(M_{BH}/10^7
M_{\odot})^{2/5} $ for the minimal peri-center distance, $ R_p=3 R_S $
(with $ R_{S}=2GM_{BH}/c^2 $ being the Schwarzschild radius of the
black hole), of the disrupted star (e.g. \citealt{Strubbe2009}), the
peak luminosity is expected to be as high as $ L_{peak} \simeq 30
L(\tau)\simeq 3 \times 10^{44} $ erg/s for $ M_{BH}=10^7 M_{\odot} $,
which is nonetheless $ \sim $5 times smaller than the Eddington
luminosity for this black hole mass.  The same calculation for $
M_{BH}=5\times 10^{6} M_\odot$ ($ \tau_{Edd}\simeq 0.19 $ yr) results
in $ L_{peak} \simeq 50 L(\tau)\simeq 5\times 10^{44} $ erg/s,
i.e. just slightly below the corresponding Eddington
luminosity. Although these estimates may be slightly
  affected by the bolometric correction, which can be estimated as $
  L_{bol}\simeq 1.1 L_{X,0.1-2.4} $ from the observed shape of the
  spectrum (a black body with $ \kappa T_{bb}=70 $ eV), and
  geometrical effects, it appears that values $ M_{BH} \lesssim
  5\times 10^{6} M_\odot$ are unlikely.

\begin{figure}
\caption{
An example of fitting the long-term X-ray light curve (the fluxes
measured during the RASS and two subsequent pointed \textit{ROSAT}
observations of RBS~1032 by a power-law decay, i.e. $ f\propto
\left(\frac{t-t_0}{\tau_{Edd}}\right)^{-5/3} $, where $ t_0 $ is the
disruption moment, for two values of the duration of the
Eddington-limited phase $ \tau_{Edd}=0.19$ and 0.25~yr (thick blue and
red lines, respectively. The corresponding peak fluxes and
luminosities (right axis, for the luminosity distance to the source $ 
d_L=114 $ Mpc) are marked by the solid blue and dashed red lines. Also
shown is the marginal detection of the source by \textit{XMM-Newton}
in late 2009, which demonstrates a strong flux drop after $\sim 20$
years. 
} 
\includegraphics[width=1.1\columnwidth]{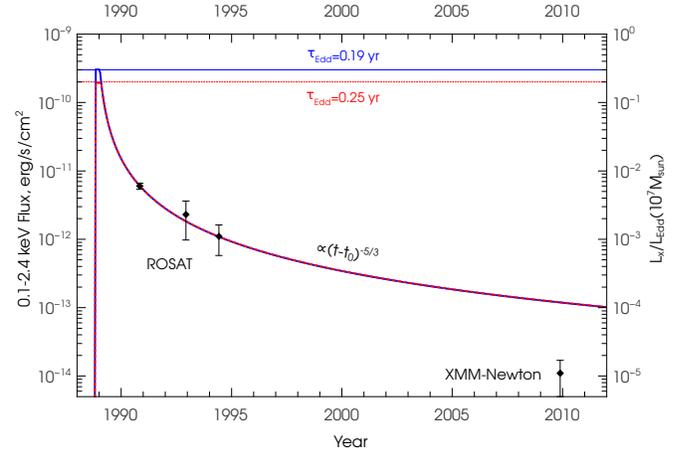}
\label{f:rbs}
\end{figure}

Therefore, the X-ray light curve seems to be consistent with a TDE
that occurred near a 'normal' SMBH with $ M_{BH}=5\times 10^6 - 10^7
M_{\odot} $, even though a simple power-law model predicts an order
of magnitude higher flux for the \textit{XMM-Newton} epoch (see
Fig.~\ref{f:rbs} for an illustration) -- this is in fact not
surprising since at this very late stage of the TDE flare the feeding
rate of the SMBH is expected to have fallen by several orders of
magnitude.   

We note that for  $ M_{BH}=5\times 10^6 - 10^7 M_{\odot} $,
one could expect a somewhat lower temperature of the multicolour
blackbody accretion disk, $ \kappa T_{bb}\simeq 40 - 50$ eV $ \propto
M_{BH}^{-1/4} $ \citep{SS1973}. However, at late phases (as suggested
here for the \textit{ROSAT} observations) when the
luminosity is as low as a few percent of the the peak value, i.e. the
mass accretion rate is a few per cent of the critical accretion
rate, the X-ray spectrum may harden due to transition to an inefficiently
cooling accretion flow and/or formation of relativistic jets. 

Therefore, 1RXS J114727.1+494302=RBS~1032 is probably a tidal
disruption event associated with a SMBH of mass $ M_{BH}\sim 5\times
10^6 - 10^7 M_{\odot} $, which was caught by \textit{ROSAT} at a
fairly late phase. We note however that the inferred black-hole mass may
be too high for the dwarf host galaxy of RBS 1032 \citep{Ghosh2006}. 

%%%%%%%%%%%%%%%%%%%%%%%%%%%%%%%%%%%%%%%%
\subsection{1RXS J130547.2+641252}

\begin{figure}
\caption{DSS image (2'x2') of the region around 1RXS J130547.2+641252}
\includegraphics[width=1.2\columnwidth]{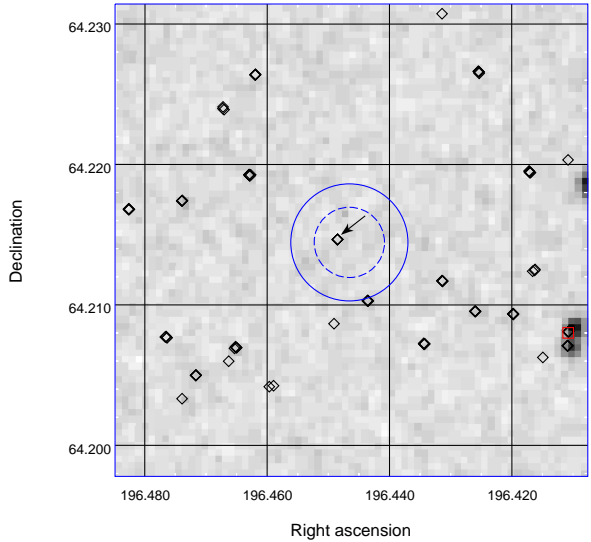}
\label{f:s8}
\end{figure}

There is only one SDSS source inside the localisation region (3 arcsec
away from the RASS-BSC centroid, see Fig.~\ref{f:s8}). It is
consistent with an extended 
source and has (22.5, 22.0, 20.9, 20.6, 20.4) $ugriz  $
magnitudes. This object is fainter than would be expected for a TDE
host galaxy, albeit not very dramatically. 

We conclude that 1RXS J130547.2+641252 is possibly a TDE. 

%%%%%%%%%%%%%%%%%%%%%%%%%%%%%%%%%%%%%%%%
 \subsection{1RXS J215101.5-302852  }

\begin{figure}
\caption{DSS image (2'x2') of the region around 1RXS J215101.5-302852. }
\includegraphics[width=1.2\columnwidth]{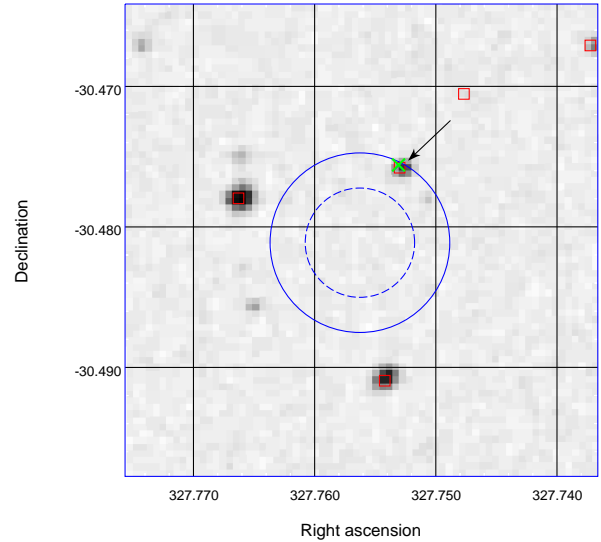}
\label{f:s10}
\end{figure}

A weak 3XMM-DR4 source (3XMM J215100.7-302832), $(4.1\pm0.8)\times
10^{-14}$~erg/s/cm$^2$ (0.2--2 keV), is detected at the boundary of
the localisation region (22 arcsec from the RASS-BSC centroid). This 
source has an optical counterpart in the DSS 
image (with $ R_F=18.6 $ for the corresponding entry in GSC-II) as
well as infrared counterparts in the 2MASS (with $ jhk $ magnitudes of 
16.849, 16.304 and 15.347) and \textit{WISE} (with $ W1, W2, W3$
magnitudes of 13.985, 12.949 and 10.192) surveys (see 
Fig.~\ref{f:s10}). The first \textit{WISE} colour, $W1-W2=1.036$, 
strongly indicates an AGN origin. 

We conclude that 1RXS J215101.5-302852 is unlikely to be a TDE. 

%%%%%%%%%%%%%%%%%%%%%%%%%%%%%%%%%%%%%%%%
\subsection{1RXS J235424.5-102053}

 {
1RXS J235424.5-102053 is located approximately 5 arcminutes from
the centre of the galaxy cluster Abell 2670. The latter enters
RASS-BSC as 1RXS J235409.4-102506, is $ \sim $ 10 times brighter than
1RXS J235424.5-102053 and has an angular extent of $ \sim $ 1.5 arcmin. 
The uncertainties in the hardness ratios for 1RXS J235424.5-102053 are
too large to draw any conclusions based on them. We note however that
the corresponding spectrum should be rather peculiar with HR1=1.0$ \pm
$0.25, i.e. having no counts in the 0.1--0.4~keV energy channel, and
HR2=-0.13$\pm $0.29 (while 1RXS J235409.4-102506 has HR1=$0.83\pm 0.06
$ and HR2=$0.24\pm0.1  $).} 

\begin{figure}
\caption{DSS image (2'x2') of the region around 1RXS J235424.5-102053 }
\includegraphics[width=1.2\columnwidth]{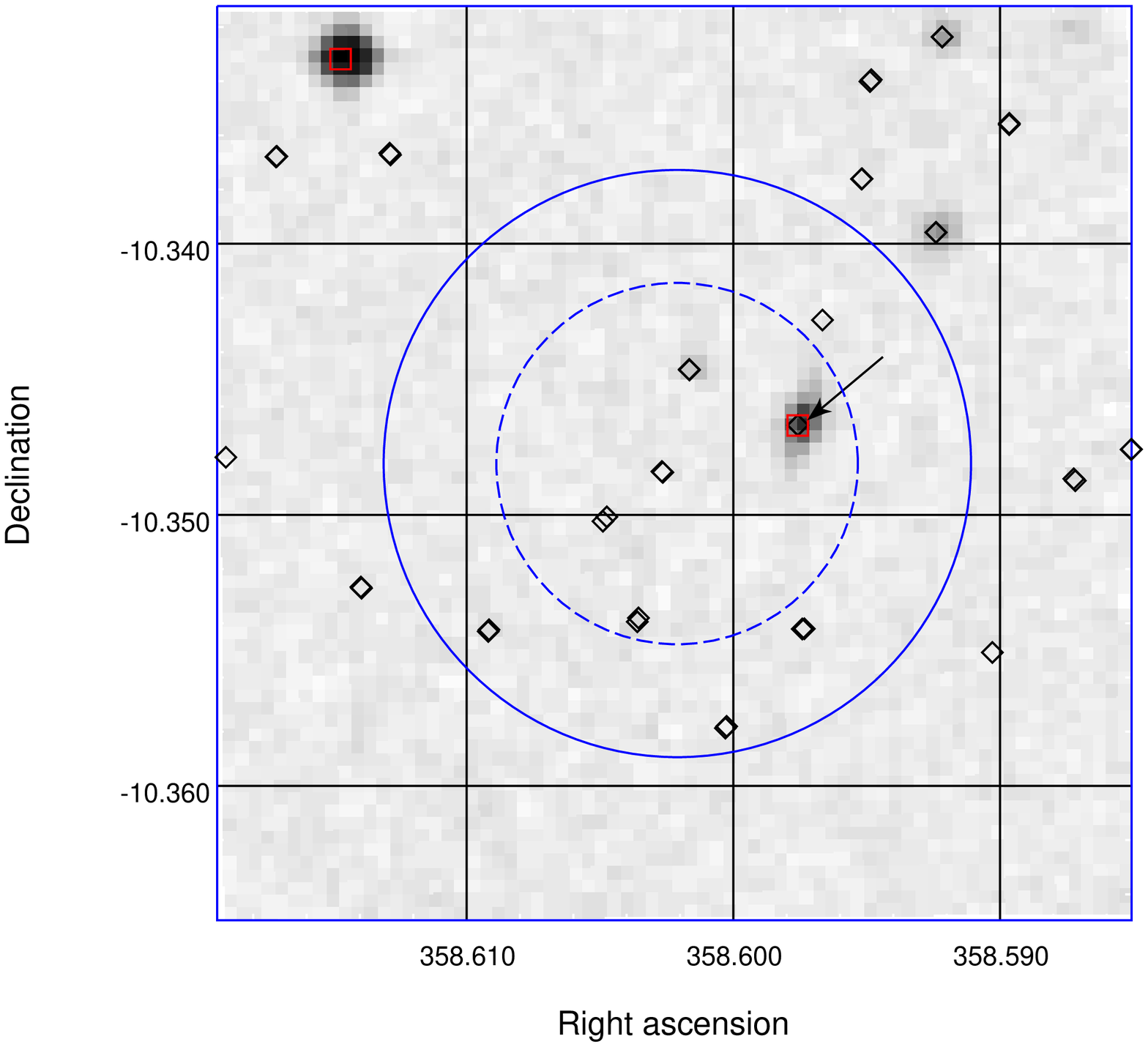}
\label{f:s11}
\end{figure}

The most prominent candidate for the counterpart of this source is a
bright SDSS galaxy (17 arcsec away from the RASS-BSC centroid, see
Fig.~\ref{f:s11}) at $ z=0.0805 $,  {which may be a
  distant member of the Abell 2670 cluster ($z=0.076$)}. Its optical
spectrum does not show signatures of luminous nuclear activity, which is
confirmed by the colours of the corresponding \textit{WISE}
counterpart ($W1-W2=0.0$). All other eight SDSS sources inside the
localisation region are faint ($ r >20 $ ) and not detected by
\textit{WISE}.  {We further note that the
    bright SDSS galaxy appears to be a disk one observed nearly edge-on,
    which might cause significant absorption of the X-ray emission in
    the softest energy band of \textit{ROSAT} and thus explain the
    measured value of HR1 close to unity.} 

 {
We conclude that 1RXS J235424.5-102053 could still be considered a
possible TDE candidate. However, we cannot exclude that it is a
spurious point-like source on a RASS image contaminated by extended
cluster emission.}

\begin{table*}
 \caption{Possible counterparts of potential TDE candidates. The coordinates of
   the most probable counterparts are marked in
   boldface.} 
\begin{tabular}{ccccccccc}\hline\hline
\smallskip
\multirow{2}*{Candidate} &\multicolumn{7}{c}{Counterpart} \\
& RA & Dec & Dist$ ^a $   & WISE, $W1$--$W2$ $ ^b $ & 2MASS, $ JHK_s $
& SDSS, Class ($ ugriz $)$ ^c $& 3XMM$ ^d $ & Nature\\\hline
\smallskip
{1RXS J002048.5-253823} &\textbf{5.2054}&\textbf{-25.6414}& 12 &13.3--12.1& 16.5--16.1--15.0&s(17.33,17.32,17.74)$ ^e $ & 1.68$ \pm $ 0.46 & AGN\\
\multirow{7}*{1RXS J005626.3-010615} 
&14.1057 & -1.1045  &13    & & &s(25.0,23.5,22.6,22.3,22.5) &&\\
& 14.1063&	-1.1053 &13    & &  &g(22.5,22.2,21.7, 22.0,21.0)& &\\
&\textbf{14.1066}	&\textbf{-1.1075}	&	16&	 16.8--15.7&  &g(22.3,21.9,21.0,20.4,20.1)&& AGN\\
&14.1045	&-1.1067	&		20&  & &	s(22.6,25.4,24.3,24.3,23.0)&&\\
&14.1093	&-1.1106	& 		23&  & & 	s(23.4,24.0,22.8,22.9,21.6)&&\\
&14.1146	&-1.1000	&		24& 	 & & g(23.9,24.3,22.5,21.4,20.9)&&\\
\smallskip
&{14.1110}&{-1.1108}&	24	&	14.8--14.6&15.8--15.1--14.9&s(22.6,21.8,20.1,18.3,17.3)& &M-star \\
\multirow{2}*{1RXS J101326.2+061202}		
&\textbf{153.3579}	&\textbf{6.1939}&	25&14.7--14.8&15.6--14.9--14.8&s(21.4,18.7,17.4,17.0,16.7)&&K-star\\
\smallskip
&153.3523&	6.2021&25&&&	s(24.6,24.0,24.1,22.6,22.1)	&&\\											
\multirow{11}*{1RXS J112312.7+012858}
&170.8046&	1.4824&7&&&		s(23.2,22.5,21.9,22.0,22.0)&&\\	
&170.8063&	1.4845&13&&&	g(22.0,22.4,21.6,20.3,19.8)&&\\	
&170.8063&	1.4859&16&&&	g(25.2,23.2,23.6,21.9,21)&&\\
&\textbf{170.8065}&	\textbf{1.4783}&21&15.1--14.8&&	g(24.1,21.6,19.9,19.2,18.8)&& TDE\\	
&170.8004& 	1.4767&24&&&	s(22.4,22.1,21.7,21.4,21.1)&&\\	
&170.8082&	1.4788&24&&&	s(24.4,22.9,21.3,20.7,20.3)&&\\	
&170.8081&	1.4781& 26&&& 	g(21.6,21.4,20.7,20.5,21.5)&&\\	
&170.8044&	1.4902&27&&&	g(24.6,25.5,23.6,21.8,22.0)&&\\	
&170.7954&	1.4818&27&&&	s(21.3,20.5,20.2,20.0,20.0)&&\\	
&170.8050&	1.4903&28&&&	s(22.2,21.3,21.0,21.4,20.8)&$6.0\pm0.8  $&AGN/CV?\\	
\smallskip
&170.8060&	1.4901&28&&&	s(25.1,25.4,25.5,21.9,23.0)&&\\	
\multirow{3}*{1RXS J114727.1+494302}
&\textbf{176.8613}&	\textbf{49.7161}&	 6 &	14.5--14.3&		15.9--15.1--15.2& g(19.1,	17.7,17.1,16.8,16.6)&&TDE(z=0.026)\\
&176.8618&	49.7143&	11&	&&						g(23.9,	23.7,22.7,23.2,22.0)&&\\
\smallskip
&176.8656&	49.7133&	15&	&&						g(22.7,	22.7,22.4,23.1,23.9)&&\\
\smallskip
{1RXS J130547.2+641252 }
&\textbf{196.4486}&	\textbf{64.2145}&	3  &&& 	g(22.5,22.0,20.9,20.5,20.4)&&TDE\\
\smallskip
 1RXS J215101.5-302852 & \textbf{327.7528}&	\textbf{-30.4758}&	22	 &14.0--13.0 &16.8--16.3--15.3 &g(18.34,18.56,17.74)$ ^e $ &$ 4.1\pm0.8 $&AGN\\
\multirow{9}*{1RXS J235424.5-102053 }
&358.6027&	-10.3484&	2&		&					&	s(24.8,	23.6,21.8,20.9,20.3)&&\\
&358.6047&	-10.3501&	12&		&					&	g(22.7,23,21.7,21.5,21.7)&&\\
&358.6017	&-10.3446&		13&		&					&s(21.4,20.8,20.9,21,21.2)&&\\
&\textbf{358.5976}	&\textbf{-10.3467}&		17			&13.7--13.7&15.4--14.7--14.3&	g(	19.6,17.7,16.8,16.4,16)&& {TDE?(z=0.081)}\\
&358.6036	&-10.3538&		21&		&					&g(23.1,21.8,20.8,20.8,20.8)&&\\
&358.5976	&-10.3542&		28&		&					&s(23,22.8,22.2,22.3,22.6)&&\\
&358.597	3&-10.3542&		28&		&					&g(23.6,22.7,22.1,22.3,22)&&\\
&358.609	3&-10.3542&		33&				17.0--16.5& & g(23.7,22.3,22,21.1,20.6)&&\\
&358.6003&	-10.3578&	36&				16.8--16.8& &	g(23.5,22.4,22.3,21,20.8)&&\\
\hline
\end{tabular}
\begin{flushleft}
\footnotesize{$ ^a $ Distance from the centroid of the RASS-BSC
  localisation, arcsec.\\  
$ ^{b} $ \textit{WISE} photometry becomes unreliable at approximately
  $W1=14$, $W2=13.5$. Values $W3>12$ (as is typical of the detections
  under consideration) should be regarded as marginal detections,
  hence we make no use of the $W3$ magnitude in our analysis (see text).\\ 
$ ^{c} $ A source type (\textit{s} for stellar-like and
  \textit{g} for galaxy-like) is given along with the $ ugriz $
  magnitudes. \\  
$ ^{d} $ The X-ray (0.2--2 keV) flux of the source in the 3XMM-DR4
  catalogue, 10$^{-14}  $ erg s$^{-1} $ cm$ ^{-2} $.\\ 
$ ^{e} $ DSS data is provided if no SDSS data is available: type
  (\textit{s} for point-like and \textit{g} for an extended source)
  and ($ B_{J} $, $ R_{F} $, $ I_{N} $) magnitudes, taken from the
  second generation Guide Star Catalog \citep{Lasker2008}. The
  characteristic uncertainty for these photometric magnitudes is
  $\approx 0.2  $ for sources at high ($ |b|>30^\circ $) galactic
  latitudes \citep{Lasker2008}.\\  
}
\end{flushleft}
 \label{t:cross}
 \end{table*}

%%%%%%%%%%%%%%%%%%%%%%%%%%%%%%%%%%%%%%%%
%%%%%%%%%%%%%%%%%%%%%%%%%%%%%%%%%%%%%%%%
\section{Discussion}
\label{s:disc}
%%%%%%%%%%%%%%%%%%%%%%%%%%%%%%%%%%%%%%%%

Since any TDEs that may be present in our sample occurred almost
25~years ago, there is not much that can be done to explore their
individual properties (see e.g. \citealt{Bower2013} who looked for the
late-time radio emission from TDE candidates discovered during the
RASS). Nonetheless, this sample is valuable for studying TDE
population properties. In particular, it allows us to put a constraint
on the rate of TDEs in the local Universe. 
%-------------------------------------------------------------------------------------------------
\subsection{Comparison with the results of Donley et al.}
\label{ss:comparison}
%-------------------------------------------------------------------------------------------------

We first compare our results with the findings of
\cite{Donley2002}. Their sample was limited to relatively bright X-ray
sources, with the unabsorbed 0.2--2.4~keV flux during the RASS higher
than $f_0= 2\times 10^{-12} $ erg/s/cm$^2 $ (assuming a power-law
spectrum with $ \Gamma=4 $), which was due to the sensitivity
threshold of the subsequent \textit{ROSAT} pointed observations
\citep{Donley2002} used in the analysis for demonstrating that the
flux of a given RASS source has undergone a strong decline. Since we
use the more sensitive \textit{XMM-Newton} observations for the same
purpose, our resulting sample goes all the way down to
the sensitivity limit of the RASS-BSC catalogue of $f= 3\times 
10^{-13} $ erg/s/cm$^2 $ (see Table \ref{t:clean}). As a result, we
can estimate the expected number of TDE candidates in our sample
from the number of candidates, $ N_0 $, found by \cite{Donley2002} as
follows:   
\begin{equation}
N\simeq N_0 \times \frac{S}{S_0}\times
\left(\frac{f}{f_0}\right)^{-3/2},
\end{equation}
where  $S/S_0  $ is the ratio of the sky areas covered by the two studies
and $ f/f_0 $ is the ratio of the corresponding limiting fluxes. This
equation is approximate since it disregards reddening of the distant
objects and assumes a simple power-law (with a slope of -3/2) $\log
N$---$\log S$ distribution, with both of these assumptions being
somewhat inaccurate (see e.g. \citealt{Khabibullin2014}). 

The sky coverage of the \textit{XMM-Newton} observations corresponding
to the 3XMM-DR4 catalogue is 794 deg$ ^2 $, i.e. almost 2\% of the
whole sky, which is 1/9th of the area covered by the \textit{ROSAT}
pointed observations \citep{Donley2002}. Given $N_0=6$
  candidates (including SBS~1620+545 with a hard X-ray spectrum and
  two candidates with AGN signatures) found by \cite{Donley2002} at $
  |b|>30^\circ $, we find from the above equation $ N=11.5 $,
   {roughly in agreement with the number (eight)} of TDE
  candidates we actually found in our study (as well as with the
  number anticipated by \citealt{Donley2002} for such a
  \textit{XMM-Newton}-RASS cross-correlation study). 

However, as we showed in Section~\ref{s:results}, three of our
candidates (1RXS J002048.5-253823, 1RXS J005626.3-010615 and 1RXS
J215101.5-302852) are likely to be high-amplitude AGN flares, and yet
another one (1RXS J101326.2+061202) might be associated with an
intense flare from a K-star.  {Finally, \textit{three}
  candidates are broadly consistent with the expectations of the TDE
  scenario. Namely, 1RXS J114727.1+494302, 1RXS J130547.2+641252 and
  1RXS J235424.5-102053 can be associated with sufficiently bright
  galaxies without signatures of luminous AGN, although we cannot
  exclude that 1RXS J235424.5-102053 is a spurious RASS source. For
  the fourth source, 1RXS J112312.7+012858, a TDE association is also
  acceptable but an AGN association could not be rejected conclusively.}

 {
Thus, among our \textit{eight} candidates, 3 or 4 could be associated
with AGN, 1 with a flaring star and 2 to 4 with TDEs. Interestingly, 
\cite{Donley2002} found a similar proportion of candidates for AGN
flares and TDEs, which supports the conclusion above that the numbers
of TDEs found in these two studies are \textit{broadly} consistent with each other.}   

%%%%%%%%%%%%%%%%%%%%%%%%%%%%%%%%%%%%%%%%
\subsection{TDE rate in the local Universe}
\label{ss:rate}

The simple estimates presented above disregard a number of important
points. First, by increasing the sensitivity of a survey, one gets the
possibility not only to detect more distant TDE flares, but also
flares at a later phase of their decay, i.e. 'older' ones at closer
distances. Second, the problem in hand involves rather soft X-ray
sources with the spectral maximum near or even below the
\textit{ROSAT} sensitivity range. This, along with the importance of
the interstellar absorption in this energy range, makes the estimates
quite sensitive to the cosmological reddening of the spectra. In
addition, the non-cubic dependence of the comoving volume on the
luminosity distance needs to be accounted for when dealing with
redshifts $ \sim 0.15 $.  { Finally, the sky coverage of
  \textit{XMM-Newton} observations is not completely independent of
  the positions of RASS-BSC sources, since some of them were the
  targets of \textit{XMM-Newton} observations.}  

 {
To estimate the last effect, we first find that $ \sim $ 8\% of all
RASS-BSC extragalactic ($|b|>30^\circ$) sources have ever been
observed by \textit{XMM-Newton}, either on purpose or serendipitously, 
whereas the sky coverage of the 3XMM-DR4 catalogue is only 2\% of that
for the RASS (essentially the whole sky). However, the relative
fraction of \textit{XMM-Newton} targets among RASS-BSC sources should 
depend on the class of object of interest.}  {As
  concerns the present study, which is aimed at non-extended, variable
  sources, the only potentially important source of bias might be
  associated with clusters of galaxies, which have often been targets of
  observations for \textit{XMM-Newton}. Indeed, since clusters contain
  a lot of galaxies, they are expected to be cites of strongly
  increased TDE activity. However, since our study probes the Universe
  out to $z\sim 0.2$, an \textit{XMM-Newton} ($\sim 15'\times 15'$)
  field of view randomly located on the sky will contain more  
  galaxies than a rich cluster, hence there should be no significant
  bias due to \textit{XMM-Newton} pointing at galaxy clusters. Moreover,
  the fact that a significant part of the XMM-DR4 
  catalogue is based on observations of fields containing galaxy
  clusters may in fact result in a decrease in the effective
  coverage for our study, since \textit{ROSAT}'s rather poor angular
  resolution prevents finding TDE flares in the vicinity of cluster cores.} 

 {
Indeed, only two out of eight candidates in our final sample of TDE
candidates, namely, 1RXS J114727.1+494302 and 1RXS J130547.2+641252,
prove to have been targets of \textit{XMM-Newton} observations, while
the others were observed serendipitously. Thus, the
\textit{XMM-Newton} sky coverage can indeed be considered almost
independent of RASS-BSC sources when dealing with TDE candidates.} 
 
Following the treatment in \cite{Khabibullin2014}, we assume now some
characteristic shape for the light curve, spectrum and energetics of a
TDE. Specifically, given $ M_{bh}=5\times 10^6 M_\odot $, one expects
a blackbody spectrum with $ \kappa T_{bb}\simeq 50 $ eV, the peak
bolometric luminosity (assumed to be equal to the Eddington
  limiting luminosity) $ L_{0}\simeq 7\times 10^{44} $ erg/s and the 
characteristic light curve time-scale $ \tau_{Edd}\simeq 0.19 $ yr
(see Sections 2.1 and 2.2 in \citealt{Khabibullin2014} and references
therein). The original spectrum will be modified by interstellar
absorption both in the host galaxy and along the line-of-sight in our
Galaxy. The median value of the Galactic HI column density for the
sources in our sample is $ N_H^0\simeq 2.7 \times 10^{20}  $cm$ ^{-2}
$ (see Table \ref{t:clean}). The absorption due to the gas in the
host galaxy could be of the same order but is poorly known (see
e.g. the comparison of the net $ N_H $ value found by the fitting the
X-ray spectra for one of our candidates, RBS 1032, to the
corresponding Galactic absorption in that direction,
\citealt{Ghosh2006}). Just for convenience, we adopt the host's
absorption column density $ N_H^{host}\simeq 2.3 \times 10^{20} $ cm$
^{-2} $, so that the net column density $ N_H\simeq 5\times 10^{20}$
cm$ ^{-2} $ (note that the Galactic absorption is
imposed in the observer's reference frame, while the host's absorption
in the rest frame of the TDE).  

In order to calculate the limiting redshift for a TDE flare at which
it could still be identified as a bright \textit{ROSAT} source
(i.e. with the count rate more than 0.05 cts/s), we performed a
\texttt{fakeit} simulation using \textit{XSPEC} \citep{Dorman2001} and
the \textit{ROSAT} response matrix as provided by the NASA's High
Energy Astrophysics Science Research Archive Center
(HEASARC)\footnote{https://heasarc.gsfc.nasa.gov/docs/rosat/rosgof.html}. 
Assuming $ M_{bh}=5\times 10^6 M_\odot $, the limiting
redshift turns out to be $ z_{lim}\simeq 0.18 $ \footnote{ {In fact, the limiting redshift depends on $ M_{bh}$ only slightly in the range from $10^6 M_\odot $ to $10^7 M_\odot $, see the corresponding discussion in \cite{Khabibullin2014}.}}. This can be
considered the characteristic depth of our sample. Using this value
and repeating the calculation presented in Section 4.1 of
\citep{Khabibullin2014}, we finally can estimate the TDE rate in the
local  Universe at  { $ \mathcal{R}\simeq 8 (4)
    \times 10^{-7}$ Mpc$ ^{-3} $ yr$ ^{-1} $, assuming that there are
    \textit{four (two)} 'true' TDEs in our sample (see Table
  \ref{t:cross} and the discussion above).} 

This estimate should still be regarded as an upper limit, since we
cannot yet conclusively identify as a TDE any of our  {four} best
candidates. Given the volume density of inactive galaxies of $2\times
10^{-2} $ Mpc$^{-3}$, our estimate translates  {to a
  rate $ R\simeq 
3\times 10^{-5} $} of TDEs yr$ ^{-1} $ per galaxy, which falls in
between the estimate of \cite{Donley2002}, $ 9.1\times 10^{-6} $ yr$
^{-1} $ per galaxy, and the one by \cite{Esquej2008}, $ 2.3\times
10^{-4} $ yr$ ^{-1} $ per galaxy. One can also try to
  predict this rate by combining equation~(29) in \cite{Wang2004}
  for the expected dependence of the TDE rate on $ M_{BH} $ and
  $\sigma $ with the well-known $ M_{BH}-\sigma $ relation 
  (e.g. \citealt{Ferrarese2005,Gultekin2009}): $ R\sim 3\times 10^{-4}
  $ yr$ ^{-1} $ per galaxy. The fact that the rates derived from the
  cited observations are all well below this prediction could be due
  to an underestimation of the mean intrinsic column density (as
  discussed by \citealt{Sembay1993}; see also the related discussions
  in \citealt{Donley2002} and \citealt{Esquej2008}), as well as due to
  uncertainties in the average properties of nuclear stellar clusters
  in nearby galaxies. 

We should also mention attempts to systematically search for TDEs
using long observations of predefined samples of galaxies, in
particular clusters of galaxies
\citep{Cappelluti2009,Maksym2010,Maksym2013}. Although such studies
have found just a few TDE candidates, their results are also 
consistent with our estimated TDE rate of a few$\times 10^{-5} $ yr$
^{-1} $ per galaxy \citep{Maksym2013}.

%%%%%%%%%%%%%%%%%%%%%%%%%%%%%%%%%%%%%%%%
\section{Conclusions}
\label{s:conc}
%%%%%%%%%%%%%%%%%%%%%%%%%%%%%%%%%%%%%%%%
 {
We presented the results of a systematic search for tidal disruption
events by looking for large (more than ten-fold) flux drops for
sources from the \textit{ROSAT} Bright Source  Catalogue during
serendipitous observations performed 10--20 years later by
\textit{XMM-Newton}. Besides a number of highly variable persistent
X-ray sources (AGN, CVs and stars), we have found up to \textit{four}
sources that could be associated with TDEs in non-active
galaxies. Specifically, \textit{three} candidates are broadly
consistent with the expectations of the TDE scenario: 1RXS
J114727.1+494302, 1RXS J130547.2+641252 and 1RXS J235424.5-102053,
although the last object may be a spurious RASS source due to
contamination by diffuse cluster emission. For 1RXS J112312.7+012858,
a TDE association is also acceptable, but an AGN origin cannot be
ruled out conclusively either. This implies the mean TDE rate $ R\sim
3\times 10^{-5} $ yr$ ^{-1} $ per galaxy within the surveyed volume,
which is broadly consistent with previous estimates.} 

Due to the high sensitivity achieved in serendipitous observations by
\textit{XMM-Newton}, our sample of TDE candidates is
significantly deeper than the previous ones by
\cite{Donley2002} and \cite{Esquej2008}. For this reason, it enables
probing the TDE rate out to a larger redshift, up to $ z\sim 0.18 $
(making some reasonable assumptions about typical TDEs). The
consistency of the TDE rate found in our work with the estimates based
on the previous studies, which probed the more local Universe,
tentatively suggests no significant evolution of the TDE rate between
$z=0$ and $z\sim 0.2$. The TDE rate derived here is also consistent  
with the rough estimate made for TDEs in clusters of galaxies
(e.g. \citealt{Maksym2013}), thus indicating no dependency on the 
large-scale environment either. 

We stress again that, in the absence of direct evidence, all of
our suggested TDEs should still be regarded as TDE candidates. Further
investigation of these sources is required (first of all, by 
means of optical spectroscopy) in order to confirm the
absence of nuclear activity and to determine the distance to the host
galaxy. Such information would help improve our estimate of the TDE
rate. In addition, distance information would make it possible to
estimate the peak luminosities of the TDEs and thus to constrain the
masses of the SMBHs in the TDE host galaxies. 

A by-product of our study is a small sample of high-amplitude AGN
flares. More detailed studies of these sources could provide some
insights into the mechanisms of AGN variability and consequently the
physics of accretion onto SMBHs (like in the case of WPVS 007,
\citealt{Grupe2013}).

\section*{Acknowledgements}
\label{s:acknowledgements}

The authors thank the anonymous referee for a number of useful
suggestions. We would also like to thank Rodion Burenin for critical
and stimulating discussions. IK acknowledges the support of the
Dynasty Foundation. The research made use of grant RFBR 13-02-01365.  

\clearpage
\newpage
\label{s:biblio}

%%%%%%%%%%%%%%%%%%%%%%%%%%%%%%%%%%%%
\end{document}